\UseRawInputEncoding
\documentclass[aps,prd, nofootinbib, notitlepage,onecolumn]{revtex4-1}
\usepackage{bm}
\usepackage{latexsym}
\usepackage{amsmath,amsfonts,amssymb}
\usepackage{graphicx,epsfig}
\usepackage{psfrag}
\usepackage{amsthm}
\usepackage{color}
\usepackage{braket}
\usepackage{indent first}
\usepackage[normalem]{ulem}
\usepackage{fancyhdr}
\usepackage{hyperref}
 \usepackage{environ}
\usepackage{mathtools}
\usepackage{float}
\def\O{\Omega}

\def\a{\alpha}
\def\b{\beta}

\def\g{\gamma}

\def\d{\delta}
\def\D{\Delta}

\def\p{\partial}
\def\f{\frac}

\newcommand{\be}{\begin{equation}}
\newcommand{\ee}{\end{equation}}
\newcommand{\bes}{\begin{equation*}}
\newcommand{\ees}{\end{equation*}}

\newcommand{\beq}{\begin{eqnarray}}
\newcommand{\eeq}{\end{eqnarray}}
\newcommand{\bseq}{\begin{subequations}}
\newcommand{\eseq}{\end{subequations}}

\NewEnviron{eqn}{
\begin{align}
\begin{split}
  \BODY
\end{split}
\end{align}
}

\NewEnviron{eqn*}{
\begin{align*}
\begin{split}
  \BODY
\end{split}
\end{align*}
}

\begin{document}
\title{ Bounds on GUP parameters from  GW150914 and GW190521}
%
%
%
%
%
%
%
%
%
%
\author{Ashmita Das}
\email{ashmita.phy@gmail.com}
\affiliation{Department of Physics, Indian Institute of Technology Madras, Chennai 600036, India}

\author{Saurya Das}
\email{saurya.das@uleth.ca}
\affiliation{Theoretical Physics Group and Quantum Alberta, Department of Physics and Astronomy, University of Lethbridge, 4401 University Drive, Lethbridge, Alberta T1K 3M4, Canada}

\author{Noor R. Mansour}
\email{noor.mansour@grad.ku.edu.kw}
\affiliation{Theoretical Physics Group, Department of Physics, Kuwait University, P.O. Box 5969, Safat 13060, Kuwait}

\author{Elias C. Vagenas}
\email{elias.vagenas@ku.edu.kw}
\affiliation{Theoretical Physics Group, Department of Physics, Kuwait University, P.O. Box 5969, Safat 13060, Kuwait}
%
%
%
%
%
\begin{abstract}
%
%
%
%
\par\noindent
We compute bounds on the GUP parameters for two versions of GUP using gravitational wave data from the events GW150914 and GW190521. 
The speed of the graviton and photon are calculated in  a curved spacetime modified by GUP, assuming that these particles have a small mass. 
The observational bound on the
difference in their speeds translates to bounds on the GUP parameters. 
These bounds are some of the best obtained so far in the context of quantum gravity phenomenology. 
%
%
%
%
\end{abstract}
\maketitle
%
%
%
%
\section{Introduction}\label{intro}
%
 %
 \par\noindent
 One of the substantial and challenging tasks in theoretical physics is to construct a consistent theory of quantum gravity (QG).
In this search, one of the promising candidates is string theory \cite{polchinski_1,polchinski_2} which postulates a unified quantum description for all the fundamental interactions, where consequently, the quantum description of gravity and all the other basic interactions emerges at an appropriate limit. 
On the other hand, Loop Quantum Gravity \cite{Smolin:2004sx,Rovelli:1990pi} and Causal Dynamical Triangulations \cite{Loll:2019rdj} (also see the references therein) dictate a direct quantization mechanism of gravity, playing no role in the unification of fundamental forces. 
Furthermore, there are schemes which have originated from the above mentioned fundamental theories, in order to exclusively study the Planck scale effects in these fundamental theories. Such schemes are the theory of modified dispersion relation (MDR) \cite{Magueijo:2001cr, Magueijo:2002am}, polymer quantization \cite{Ashtekar:2002sn,Ashtekar:2002vh,Date:2011bg}, and doubly special relativity (DSR) theories  \cite{Magueijo:2001cr, AmelinoCamelia:2000mn, AmelinoCamelia:2000ge, Magueijo:2002am, Cortes:2004qn}.

Therefore,  we have some promising theories of QG and at the same time we must be ascertained with the observational signatures of QG. It is worth to be mentioned that several laboratory-based experiments have been proposed in order to search for the QG signatures \cite{Pikovski:2011zk, Marin:2013pga, Bawaj:2014cda}. 
%
%
%
%
However, there has not been any experimental or observational support for any theory of QG so far. 
Thus, it is important to explore potential signatures of these theories in the light of current or future experiments. This has been the subject of study of Quantum Gravity Phenomenology \cite{AmelinoCamelia:2008qg, Hossenfelder:2012jw}. 
%
%

The theories of QG 
predict the existence of a minimum measurable length scale 
${\cal O}(\ell_{Pl})$ \cite{Amati:1988tn,Garay:1994en,Kempf:1994su}, 
which constrains the measuring device from 
probing an arbitrarily small length scale. 
This in turn implies the modification of the Heisenberg Uncertainty Principle (HUP) to the Generalized Uncertainty Principle (GUP), which has played an important role in the development of QG phenomenology
\cite{Maggiore:1993rv,Maggiore:1993zu,Maggiore:1993kv,Scardigli:1999jh}.
%
%
Its implications have been explored in the context of Hawking radiation from a black hole (BH) spacetime \cite{Adler:2001vs, Nozari:2008gp, Alonso-Serrano:2018ycq}, BH thermodynamics \cite{Myung:2006qr,Bargueno:2015tea,Gangopadhyay:2018hhw}, Friedman-Robertson-Walker (FRW) cosmology, FRW thermodynamics 
\cite{Ali:2014hma,Ali:2015ola}, condensed matter systems
 \cite{Das:2011tq}, neutrino oscillations \cite{Sprenger:2010dg},  other quantum mechanical systems
\cite{Ali:2011fa} etc. 
%
%
%
GUP also gives rise to a modified energy momentum dispersion relation $E(p)$ which has potential observational consequences, as shown by a number of authors
\cite{Magueijo:2002am,Gambini:1998it,Alfaro:1999wd,AmelinoCamelia:2001dy,AmelinoCamelia:2004fk,AmelinoCamelia:2005ik,Majhi:2013koa}.

\par

In case of the theories with MDR, the authors in 
Refs.\cite{Ellis:2016rrr,Arzano:2016twc,Gwak:2016wmg} have shown that this results in the difference between the speed of gravitational waves (GW) and electromagnetic waves from the same astrophysical event. 
This is compatible with data from the GW event GW150914, as reported by the Laser Interferometer Gravitational-Wave Observatory (LIGO) Scientific and Virgo Collaborations \cite{Abbott:2016blz,TheLIGOScientific:2016src}. 
Ref. \cite{Arzano:2016twc} also reported an observationally  compatible version of MDR. 
These potential departures from the standard dispersion relation
show that it is important to pursue 
the search for the QG signatures in GW data, which may shed  light on QG theories.

Returning to the issue of GUP, we mention that the authors in Refs. \cite{Feng:2016tyt, Roushan:2019miz} have constrained the parametric space of several GUPs and obtained the upper bounds on the corresponding GUP parameters (defined in  Section \ref{gup_description}). 
In particular, in Ref. \cite{Feng:2016tyt} the observational data of GW150914 were used while in Ref. \cite{Roushan:2019miz} the observational data of GW170814 were used. Both groups have followed a similar approach to constrain the GUP parameters by calculating the difference between the speed of GWs, i.e., gravitons, and that of the light waves, i.e., photons. 

\par
In the aforementioned papers there are two limitations. The first  limitation is due to the fact that the photon speed is considered  unmodified (constant). However, the speed of the photon can be modified due to  QG effects when it has a small mass.
The second limitation of the aforementioned papers is that their analysis was done in flat spacetimes, although the GWs from the events GW150914 and GW170814 originated in a spacetime region of high curvature, from the merging of the two stellar mass BHs \cite{Abbott:2016blz,TheLIGOScientific:2016src,TheLIGOScientific:2017qsa}. 
Although spacetime is almost flat at the site of the detectors and, hence, their results, e.g. those of Refs. \cite{Feng:2016tyt, Roushan:2019miz}, should be valid at least approximately. 
Strictly speaking, the analysis should be carried out in a curved background spacetime, if at least to simply establish the limits of validity of the earlier results. Moreover, curvature effects may indeed be important for future events (those from regions of even higher curvatures) and for future detectors, with even higher accuracies.

Motivated by the above, in the present manuscript, we consider a curved spacetime, reasonably approximated by the 
Schwarzschild black hole metric at large distances, 
and obtain the modifications in the velocity
\footnote{In this work, we use the terms ``velocity" and ``speed" interchangeably.} of the gravitons and photons due to the GUP effects.
We focus on two specific GUPs. The first, 
 proposed by Kempf {\it et al.} in Ref. \cite{Kempf:1994su}, while the second form is proposed in Refs \cite{Das:2008kaa, Das:2010sj,Das:2009hs,Ali:2009zq, Ali:2010yn}, known as the LQGUP  since in this version there are both linear and  quadratic terms in momentum. 


\par
The remainder of this paper is organised as follows. In Section \ref{gup_description}, we briefly describe the two GUP models.
In Section \ref{bound_sch} we calculate the difference in the speed of gravitons and photons in the Schwarzschild background without GUP effects where  the effect of the curved background is expressed through the metric element $g_{00}(r)$ of the Schwarzschild black hole spacetime. 
In the limit of $r\rightarrow\infty$, our result reproduces the one for flat spacetime, as expected. 
Moreover, utilizing the data from GW150914, we get the upper bound for the difference in the speed of the gravitons and photons. 
This upper bound will be the reference point in the next sections in which the QG effects will be taken into consideration. 
In Section \ref{gup_quadrat}, we obtain the difference in the speed of gravitons and photons in a Schwarzschild black hole background including the GUP effects, which are expressed through the terms involving the dimensionless GUP parameter $\beta_{0}$ (defined in Section \ref{gup_description}). 
Utilizing the data from GW150914, we bound the difference in speeds and, thus, we get an upper bound for $\beta_{0}$. 
In addition, we consider the case in which the photon speed is GUP-modified as  is the graviton speed. 
In this case, in order to get an upper bound, we use the data from GW190521  \cite{Abbott:2020tfl}, since for this GW event, we have apart from the data from LIGO Scientific and Virgo Collaborations, its  electromagnetic counterpart from the data of the Zwicky Transient Facility (ZTF) \cite{Graham:2020gwr}. In Section \ref{gup_linear}, we follow exactly the same analysis with that in Section \ref{gup_quadrat}, but for the case of the LQGUP version and, thus, we obtain the upper bound on the dimensionless LQGUP parameter $\alpha_{0}$ (defined in Section \ref{gup_description}) which is the lowest in the literature.
In Section \ref{conclusions}, we conclude and present our results. Finally, we provide three appendices in which we have reconsidered all the previous three sections, namely \ref{bound_sch}-\ref{gup_linear}, when the very speculative case of massive photons and gravitons is included without using the small mass approximation.
\section{Brief description of GUP models}\label{gup_description}
In this section firstly, we focus on the GUP model as proposed by Kempf {\it et al.} \cite{Kempf:1994su}, which consists of the following commutator and the corresponding resulting GUP
\bseq
\begin{align}
\big[\bf{x},\bf{p}\big]=\, i\,\hbar\,\big(1+\,\b\,\bf{p}^2 \big)
\label{gup1_commutation}\\[10pt]
\D x\,\D p\,\geq\,\f{\hbar}{2}\bigg[\,1+\,\b (\D p)^2\,\bigg]
\label{gup1_uncertainty}
\end{align}
\eseq
where ($\bf{x},\,\bf{p}$) are the 3-position and 3-momentum operators and $\D x$ and $\D p$ are their uncertainties, respectively. The parameter $\b$ is given as $\b=\f{\b_{0}\ell_{Pl}^{2}}{\hbar^2}=\f{\b_0}{M_{Pl}^{2}c^2}> 0$ and is independent of $\D x,\,\D p$ while $\b_0$ is the dimensionless GUP parameter. 
In the above, $M_{Pl}$ is the Planck mass, such that
$M_{Pl}\,c^2 \sim 10^{16}$ TeV.
As is well known, Eqs.(\ref{gup1_commutation}) and (\ref{gup1_uncertainty}) imply a minimum  position uncertainty: $\D x_0=\,\hbar\sqrt{\b}=\sqrt{\b_0}\, \ell_{Pl}$. 
%
%
%
%
\par\noindent
\\
Secondly, we consider the generalization proposed in Refs.  \cite{Das:2008kaa, Das:2010sj,Das:2009hs,Ali:2009zq, Ali:2010yn},
where the GUP modified position and momentum obeys the following commutation relation, 
%
%
\beq
[x_i,\,p_j]=i\hbar\bigg[\d_{ij}-&\a\bigg(p\,\d_{ij}+\f{p_ip_j}{p}\bigg)+\a^2\bigg(p^2\,\d_{ij}+3\,p_ip_j\bigg)\bigg].
~~~~~~~~~~~~~~
\label{nrgup_1_a}
\eeq
In the above, $i,j=1,2,3$,  
$p^2=\sum_{j=1}^{3}\,p_{j}p_{j}$, and  $\a=\a_0/M_{Pl}c=\a_0\ell_{Pl}/\hbar$.
In this case, $\a_0$ is a dimensionless parameter
which will henceforth be referred to as the LQGUP parameter. 
We are motivated to work with the present LQGUP model due to the following reasons. First, it can be shown that  Eq. (\ref{nrgup_1_a}) is consistent with DSR theories. Second, this LQGUP model can be proposed from a purely phenomenological point of view.  
Third, generally, one could have a nonlinear model of GUP with all the powers of GUP parameters which supports the inclusion of linear term along with the quadratic one.
Finally, it has also been shown that the LQGUP model can be viewed as an ``effective theory" from a fully Lorentz covariant theory as described in Ref. \cite{bossodas}.
This LQGUP model gives rise to the following modified position-momentum uncertainty relation 
\beq
\D x\,\D p &\geqslant&\,\f{\hbar}{2}\bigg[1-2\a\braket{p}+4\a^2\braket{p^2}\bigg]\nonumber\\
&\geqslant&\,\f{\hbar}{2}\bigg[1+\bigg(\f{\a}{\sqrt{\braket{p^2}}}+4\a^2\bigg)\,\D p^2+4\a^2\braket{p}^2-2\a\sqrt{\braket{p^2}}\bigg]~.
\label{nrgup_2}
\eeq
%
%
The modified commutation relation, namely Eq. (\ref{nrgup_1_a}), and the modified uncertainty principle, namely Eq. (\ref{nrgup_2}), imply a minimum measurable length and a maximum measurable momentum of the form
\beq
\D x\,\geqslant\,(\D x)_{{\rm min}}\, \approx\, \a_0\,\ell_{Pl},\,\,\,\,\,\,\,\,\,\,\,\,\,\,\,\,\,\,\,\,\,\,\,\,\,\,\, \D p\,\leqslant\,(\D p)_{{\rm max}}\,\approx \,\f{M_{Pl}\,c}{\a_0}~.
\label{minimal_gup2}
\eeq

%
%
%
%
\par\noindent
Note that for both GUPs under consideration, one assumes 
$[p_i,\,p_j]=[x_i,\,x_j]=0$. 
%
%
The dimensionless GUP parameters $\b_0$ and $\a_0$
are sometimes assumed to be ${\cal O}(1)$
\footnote{Recently, in Ref. \cite{Scardigli:2016pjs} and utilizing the GUP version, a numerical value of the dimensionless GUP parameter, namely  $\b_0$, was obtained, namely $82\pi/5$. 
Furthermore, in Ref. \cite{Vagenas:2018zoz} and utilizing the LQGUP version, the dimensionless GUP parameter, namely $\a_0$, was shown to be proportional to powers of the dimensionless ratio ($M_{Pl}/M)$.}.  
However, in this work, we will {\it not} make this assumption
{\it a priori}. We keep them arbitrary, examine the consequences, and let experiments and observations decide on their values. 
%
%
We note however, that the above parameters give rise to the 
length scales $\a_0\,\ell_{Pl}$ and $\sqrt{\beta_0}\,\ell_{Pl}$. Assuming that the
length scales are no smaller than the Planck length 
$\ell_{pl} \approx 10^{-35}$ m (as the physics beyond that scale is completely unknown) and do not exceed the electroweak scale, which is about $10^{-18}$ m, one gets a set of natural bounds $1 \leq \alpha_0 \leq 10^{17}$ and 
$1 \leq \beta_0 \leq 10^{34}$.
The work of various authors,  e.g. Ref. \cite{Ali:2011fa},
and their exploration of low energy effects on GUP set stricter bounds on the parameters, and lend further credence to the possible existence of an intermediate scale between the Planck and electroweak scales. 

\par
\section{Speed of graviton and photon in the Schwarzschild background}\label{bound_sch}
%
%
%
%
%
%
%
\par\noindent
As mentioned in the introduction, we model a generic curved spacetime at large distances from the source by a 
4-dimensional Schwarzschild metric 
\beq
ds^2=-f(r)c^{2}dt^2+\frac{1}{f(r)}dr^2+r^2\,d\Omega^{2}
\label{sch_1}
\eeq
where  $f(r)=\bigg(1-\frac{2\,GM}{c^2\,r}\bigg)$ and
$d\O^2=\,r^2\,d\theta^2+r^2\,{\rm sin}^2\theta\,d\phi^{2}$. 
We write the squared $4$-momentum of a particle of mass $m$ in the aforesaid background as 
\beq
p_{\mu}\,p^{\mu}\,=g_{00}(p^{0})^2+\,\underbrace{g_{ij}p^{i}p^{j}}_{p^2}
\label{sq_mom_1}
\eeq
%
%
%
%
%
%
%
%
%
The standard dispersion relation dictates that $p_{\mu}p^{\mu}=-\,m^2c^2$ 
and, therefore, the above equation becomes
\beq
(p^{0})^2=\,\f{1}{g_{00}}\,\bigg(-p^2\,-m^2c^2\bigg)~.
\label{zero_mom_1}
\eeq
As usual, one defines the energy of a particle in this background using the timelike Killing vector field $\xi_{0}$ in the Schwarzschild background 
\beq
\f{E}{c}=\,-\xi_a\,p^a=\,-\,g_{ab}\,\xi^a\,p^b
\label{energy_1}
\eeq
where 
$\xi^{a}=(1,0,0,0)$ is the timelike Killing vector. 
Therefore, from Eq. (\ref{energy_1}) we get
\beq
E=\,-g_{00}\,c p^{0}~.
\label{energy_2}
\eeq
%
%
%
%
We now specify the particle as graviton with its rest mass and energy to be  $m=m_g$ and $E=E_g$, respectively. From Eq. (\ref{zero_mom_1}), we get the energy of the graviton to be
\beq
E_{g}=\,\sqrt{-\,g_{00}}\,(p^2c^2+\,m_{g}^{2}c^4)^{1/2}~.
\label{energy_3}
\eeq
It may be noted that extreme cosmic phenomena cause the fluctuations in spacetime, which in turn produce gravitons. 
These can then propagate as a GW and hit 
the GW detectors such as the LIGO and Virgo detectors. 
The speed of gravitons can be calculated by using the group velocity of the wave front
\beq
v_g=\f{1}{\sqrt{-\,g_{00}}}\,\f{\p E_g}{\p p}~.
\eeq
Using Eq. (\ref{energy_3}), we obtain
%
%
%
%
%
%
%
%
%
%
%
%
\beq
v_g=\,c\bigg[1-\f{m_{g}^{2}c^4}{E_{g}^{2}}\,\bigg(1-\f{2GM}{r\,c^2}\bigg)\bigg]^{1/2}~. 
\label{vel_gr_1}
\eeq
Expanding the term within the square brackets in Eq. (\ref{vel_gr_1}) yields
\beq
v_g =\,c\bigg[1-\f{m_{g}^{2}c^4}{2\,E_{g}^{2}}\,\bigg(1-\f{2GM}{r\,c^2}\bigg)-\f{1}{8}\,\f{m_{g}^{4}c^8}{E_{g}^{4}}\,\bigg(1-\f{4GM}{r\,c^2}\bigg)-\mathcal{O}(G^2)\bigg]~.
\label{vel_gr_2}
\eeq
Since $\ell_{Pl}=\sqrt{G\hbar/c^3}$ implies that ${\cal O} (G^{2})\sim{\cal O} (\ell_{Pl}^4)$, 
one can neglect these higher order terms in the above equation.
%
%
%
%
%
%
%
%
%
At this point a number of comments are in order. First, it is evident that $\D v_g = v_g - c \neq 0$ only when the gravitons are massive, i.e., $m_g\neq 0$. 
Second, by the same token, the speed of light remains intact as long as the photon is considered massless. 
%
%
%
%
Third, in Refs. \cite{Abbott:2016blz,TheLIGOScientific:2016src} the signal of the event GW $150914$ is peaked at $\nu = 150$ Hz, which leads to the maximum energy of gravitons to be $E_g = h\nu \approx 6.21 \times 10^{-13}$ eV. In addition, we know that if  the dispersion relation for the GW is modified, the upper bound on the Compton wavelength is constrained to be  $\lambda_g > 10^{16}$ m  and, thus, one obtains an upper bound for the mass of gravitons to be $m_g \leq 1.24 \times 10^{-22}\,{\rm eV}/c^2$ \cite{Will:1997bb}  
\footnote{ In  different contexts, one can propose even smaller bounds on the graviton mass, e.g.  $m_g \leq 10^{-32}\,{\rm eV}/c^2$ \cite{ahmeddas}.}.
With these values, the term  $\f{m_{g}^{2}c^5}{2E_{g}^{2}}$ in Eq. (\ref{vel_gr_2})
becomes $\f{m_{g}^{2}c^5}{2E_{g}^{2}} =
5.98\times 10^{-12}$ m/sec \cite{Feng:2016tyt}.
Therefore, one can neglect terms of higher order in  $\f{m_{g}^{2}c^5}{2E_{g}^{2}}$ and the speed of graviton from
Eq. (\ref{vel_gr_2}) is
%
%
%
%
%
\beq
v_g=\,c\bigg[1-\f{m_{g}^{2}c^4}{2E_{g}^{2}}\,\bigg(1-\f{2GM}{rc^2}\bigg)\bigg]~.
\label{vel_gr_4}
\eeq
%
%
%
%
%
%
%
%
%
We estimate the factor $\f{2GM}{rc^2} \sim 0.144\times 10^{-19}$ by using the data from the event GW150914 as detected by LIGO Scientific and VIRGO Collaborations, where the mass and the luminosity distance of the BH are recorded to be $M=62\, \mbox{M}\odot$ and $r=410$ Mpc, respectively \cite{Abbott:2016blz,TheLIGOScientific:2016src}. Therefore, one can practically neglect the curvature effect of the spacetime and obtain the difference between the speed of graviton and photon as follows 
%
%
%
\beq
\D v_g  = \,c-v_g\, = \f{m_{g}^{2}c^5}{2E_{g}^{2}}
\label{diff_v_gr}
\eeq
which agrees with the bound obtained in Ref. \cite{Feng:2016tyt} in  which a flat spacetime was considered.
%
%
%
%
%
%
%
%
%
%
%
%
%
Finally, utilizing the bound on the graviton mass, one gets the bound on the difference between the graviton and photon as
\beq
\D v_g  \leq  5.98\times 10^{-12} \mbox{m/sec}~.
\label{bound}
\eeq
\par\noindent
At this point, it is noteworthy that 
in our present work, we obtain the bounds on the GUP parameters  by utilizing the data from the events GW150914 and GW190521 as detected by LIGO Scientific and VIRGO Collaborations. The detected gravitational waves  of these GW events are produced from the merging of rotating binary black holes. For instance, in the event GW150914 the final spin $a_s=\,\f{c J}{G M^2}$ of the BH has been detected as $\sim 0.7$, while the luminosity distance of the BH is $r = 410$ Mpc and  $J$ depicts the final angular momentum of the BH. 
The parameter $a$ which appears in the Kerr metric is defined to be $a=\f{J}{M\,c}$  and, thus, one obtains $a_s=\,\f{c^2\,a}{GM}$ which in turn produces $\f{a}{r}=\f{G\,M}{c^2}\,\f{a_s}{r}$.
Using $a_s=0.7,\,r=410$ Mpc, we get $a/r\ll\, 1$. 
Now we write the Kerr metric in Kerr-Schild cartesian coordinates 
\cite{Visser:2007fj}
%
%
%
%
\beq
ds^2=\,-dt^2+dx^2+dy^2+dz^2+\f{2Mr^3}{r^4+\,a^2 z^2}\,\bigg[dt+\f{r(x dx + y dy)}{a^2+\,r^2}+\f{a(y dx- x dy)}{a^2+\,r^2}+\f{z}{r} dz\bigg]^2
\label{ks_1}~.
\eeq
%
%
%
Following the above analysis, it is completely reasonable to 
implement $a/r \ll 1$ in Eq. (\ref{ks_1}) and employing the advanced EddingtonÐFinkelstein (EF) coordinates, one obtains Eq. (\ref{ks_1}) as the Schwarzschild metric in EF coordinates
\footnote{Note that as similar to the term $a/r=(GM/rc^2)a_s$, the term $(2GM)/rc^2$ (which appears in the Schwarzschild metric), is also $<< 1$. However it can be shown that $a/r=(GM/rc^2)a_s$ is more suppressed than $(2 G M)/rc^2$, which justifies neglecting $a/r$ in comparison with $(2GM)/rc^2$. Thus, $(2GM)/rc^2$ becomes the leading contributing factor from the curvature of  spacetime no matter how tiny is the contribution. Therefore, in the context of this work, where our prime motivation is to work in the background of curved spacetime, we keep this term in our analysis despite of its tiny contribution in the bound of the GUP parameters, and ignore the $a/r$ term.}.
Therefore, it is legitimate to work with the Schwarzschild spacetime instead of the Kerr spacetime in order to explore the phenomenology involved in our present work.
\section{Bound on the GUP parameter from GW150914}
\label{gup_quadrat}
%
%
%
%
%
\par\noindent
In this section, we will study the modifications in the difference between the speed of graviton and photon, i.e., $\D v_g$, in Schwarzschild spacetime while incorporating the GUP defined by 
Eqs. (\ref{gup1_commutation}) and
(\ref{gup1_uncertainty}). 
For this reason, the following variables are defined
\bseq
\begin{align}
&x_i=x_{0i}\,,\,\,\,\,\,\,\,\,\,\,\,\,\,\,\,\,\,\,\,p_0=k_0
\label{gupmodel_1}\\[10pt]
& p_i=\,k_i\,(1+\b\,k^2) \label{gupmodel_1_a}
\end{align}
\eseq
where $x$ and $p$ are the physical position and momenta, while $x_0$ and $k$ are auxiliary ``canonical variables", such that 
$[x_{0i}, k_{j}] = i\hbar\, \delta_{ij}$ .
%
Next, we expand the 
squared 4-momentum, using Eq. (\ref{gupmodel_1_a}) as follows
\beq
p_{\a}p^{\a}&=&\,g_{00}(p^{0})^{2}+p^2\nonumber\\[10pt]
&=&\,g_{00}(p^{0})^{2}+\,k^2[1+2\b\,k^2+\,{\cal O}(\b^2)]~.
\eeq
It is easily seen that the last term is of ${\cal O}(\ell_{Pl}^4)$ and, thus, it can be ignored compared to the linear term in $\b$. In addition, the physical 4-momentum does not satisfy the standard dispersion relation, namely $p_{\a}p^{\a}\,\neq \,-m^2c^2$, while the non-GUP-modified 4-momentum satisfies the standard dispersion relation, namely $k_{\a}k^{\a}\,= \,-m_g^2c^2$. 
Therefore, employing the above equation, one ends up with
\beq
g_{00}(p^{0})^{2}=\,-m_{g}^{2}c^2-p^2\,+2\b\,k^2k^2~.
\label{mom_gr_1}
\eeq
\par\noindent
Now one can take the inverse transformation of Eq. (\ref{gupmodel_1_a}) and write $k$ as a function of  $p$ in the form
\beq
k^2=\,p^2\,(1-2\b\,p^2)~.
\label{gupmodel_1_b}
\eeq
%
%
%
Next we substitute Eq. (\ref{gupmodel_1_b}) into Eq. (\ref{mom_gr_1}) and we expand in terms of the GUP parameter. Then, due to the smallness of the QG corrections, we ignore the higher order terms in $\beta$, and, thus, we get
%
%
\beq
(p^{0})^{2}=\,\f{(-1)}{g_{00}}\bigg[\,m_{g}^{2}c^2+\,p^2(1-2\b\,p^2)\,\bigg]~.
\label{mom_gr_2}
\eeq
\par\noindent
Employing Eq. (\ref{energy_2}), the energy of the graviton becomes  of the form
\beq
E_g=\,\sqrt{-g_{00}}\, \bigg[m_{g}^2c^4+\,p^2c^2\,(1-2\b\,p^2)\bigg]^{1/2}~.
\label{energy_gr_1}
\eeq
The above equation now implies the following  
GUP-modified group velocity of the graviton
\beq
\tilde{v}_g=\,\f{1}{\sqrt{-\,g_{00}}}\,\f{\p E_g}{\p p}=\,\bigg[m_{g}^{2}c^4+\,p^2c^2\,(1-2\b\,p^2)\bigg]^{-1/2}\,\bigg[pc^2\,(1-2\b\,p^2)-\,2\b\,pp^2c^2\bigg]~.
\label{vel_gup_1}
\eeq
Recalling the graviton mass bound $m_g\,\leq 1.24 \times 10^{-22}\,{\rm eV}/c^2$, 
we neglect the graviton mass term in comparison with 
the remaining terms in Eq. (\ref{vel_gup_1}) and obtain
%
%
%
%
%
\beq
\tilde{v}_g=\,c\,(1-3\b\,p^2)~.
\label{vel_gup_2}
\eeq
At this point, we need to express the speed of the graviton, i.e., $\tilde{v}_g$, in terms of $E_g$, thus we implement the iteration method. First, we get the zeroth order solution for $p$ by setting $\b =0$ in Eq. (\ref{energy_gr_1}) which leads to $p=\bigg[\f{E_{g}^{2}}{(-g_{00})c^2}-m_{g}^{2}c^2\bigg]^{1/2}$. Then,
substituting the zeroth order solution in Eq. (\ref{energy_gr_1}), 
employing $g_{00}=\,-\,\bigg(1-\f{2GM}{r\,c^2}\bigg)$, and adopting the small
$m_g$ approximation \footnote{A full analysis, with no small $m_g$ approximation, is given in Appendix B.}, we finally obtain  $p$ as a function of $E_g$ \footnote{For a more detailed derivation see Appendix B.}. So, the speed of the graviton will read %
\beq
\tilde{v}_g&=&\,c\,\bigg[1-\f{3\b\,E_g^2}{c^2}\,\bigg(1-\f{2GM}{r\,c^2}\bigg)^{-1}\,\bigg]\nonumber\\
&=&\,\,c\,\bigg[1-\f{3\b\,E_g^2}{c^2}\,\bigg(1+\f{2GM}{r\,c^2}\bigg)\bigg]
\label{vel_gup_3}
\eeq
\par\noindent
where we have neglected higher order terms ${\cal O} (G^{2})\sim{\cal O} (\ell_{Pl}^4)$.
\par\noindent
Therefore, for the GUP under consideration, 
the difference between the speed of the graviton and photon, i.e., $\D \tilde{v}_g$, is given by  
\beq
\D \tilde{v}_{g}=\, c-\,\tilde{v}_g=\,\f{3\b\,E_g^2}{c}\,\bigg(1+\f{2GM}{r\,c^2}\bigg)~.
\label{dif_v_gup_1}
\eeq
%
%
%
%
%
%
%
%
%
In Section II, we derived a bound (see Eq. (\ref{bound})) for the difference in the speed of photon and graviton, which yields
\beq
\D\tilde{v}_g&=&\f{3\b\,E^2}{c}\,\bigg(1+\f{2GM}{r\,c^2}\bigg)\,\leq\,\D v_g\nonumber\\
&\implies\,& \b_0\,\leq\,\f{\D v_g\,M_{Pl}^{2}\,c^3}{3\,E_{g}^{2}}\,\bigg(1-\f{2GM}{r\,c^2}\bigg)~.
\label{bound_gup_1}
\eeq
\par\noindent
At this point, we note a number of comments. First, following the discussion below Eq. (\ref{vel_gr_4}), we state that the contribution from flat spacetime dominates over that coming from the curvature effect of the spacetime. This  indicates that with respect to the current observational data the upper bound on $\b_0$ in the curved spacetime, is left unmodified from that of the flat spacetime result \cite{Feng:2016tyt}. 
Second, taking the $r\,\to \infty$ limit in Eq. (\ref{dif_v_gup_1}), 
one reproduces the flat space result, i.e., $\D \tilde{v}_g=\,\f{3\b_0\,E_g^2}{M_{Pl}^{2}\,c^3}$, as expected \cite{Feng:2016tyt}.
Third, for the event GW150914, we take the final black hole mass to be $M=62\, \mbox{M}\odot$ and the luminosity distance to be $r=410$ Mpc. Substituting  the aforesaid values in Eq. (\ref{bound_gup_1}) and using the upper bound from Eq. (\ref{bound}), i.e. $\D v_g  =  5.98\times 10^{-12} \mbox{m/sec}$, we obtain
\be
\b_0\,\leq 2.56\times 10^{60}~.
\ee
\par\noindent
It should be noted that this bound is in agreement with the corresponding one obtained in Ref. \cite{Feng:2016tyt}. In addition, it is one of lowest bounds  among those obtained from observations in the sky \cite{Roushan:2019miz,Scardigli:2014qka} and expected to improve significantly over time with increasing accuracies of GW observations.
%
%
%
%
\subsection{$\D \tilde{v}_g$ with the GUP-modified speed of photon and GW190521}
%
%
%
%
%
\par\noindent
One now considers the case in which  the velocity of the photon is also GUP-modified similar to the velocity of the graviton. Therefore, adopting the previous analysis for the velocity of the graviton with the small mass approximation, the velocity of the photon takes the form (see Eq. (\ref{vel_gup_3}))
\beq
\tilde{v}_{\g}=\,c\,\bigg[1-\f{3\b\,E_{\g}^{2}}{c^2}\,\bigg(1+\f{2GM}{r\,c^2}\bigg)\bigg]
\label{vel_photon_1}
\eeq
where $\tilde{v}_{\g}$ is the GUP-modified velocity of the photon and $E_{\g}$ is the energy of the photon. Hence,   the difference in the speeds, i.e., $\D \tilde{v}_{g}$, reads now 
\beq
\D \tilde{v}_g=\,
\left| \tilde{v}_{\g}-\tilde{v}_g \right|
=\,
\f{3\b}{c}\,
\left|\big(E_{g}^{2}-E_{\g}^{2}\big)\right|\,\bigg(1+\f{2GM}{r\,c^2}\bigg)~.
\label{diff_photon_1}
\eeq
At this point, it should be stressed that since the speed of the photon may vary due to the curvature and QG effects, the speed of the graviton may become larger than the modified speed of photon. Therefore, from now on, we will take the absolute value of the difference between the speed of the graviton and the speed of the photon in order to define the quantity $\D \tilde{v}_g$.
Finally, employing the bound on the graviton mass in  Eq. (\ref{diff_photon_1}), we obtain a bound on the GUP parameter $\b_{0}$ 
\beq
\b_{0}\,\leq\,\f{\D v_g\,M_{Pl}^{2}\,c^3}{3\,\left|\big(E_{g}^{2}-E_{\g}^{2}\big)\right|}\,\bigg(1-\f{2GM}{r\,c^2}\bigg)~.
\label{bound_photon_1}
\eeq
A number of comments are now in order.
First, in order to compute the bound 
given in Eq. (\ref{bound_photon_1}), one needs to use data from a GW event for which the GW signal and its EM counterpart have both been detected. Until now, the only such GW event is  the GW190521, a result from the merger of two black holes \cite{Abbott:2020tfl}. 
Second, for the event GW190521, we take the final black hole mass to be $M=142\, \mbox{M}\odot$ and the luminosity distance to be $r=5.3$ Gpc. 
Thus, one can estimate the factor $\f{2GM}{r\,c^2}$ to be approximately equal to $0.025\times 10^{-19}$ and neglect the curvature effect of the spacetime while determining the upper bound on $\b_0$ from the Eq. (\ref{bound_photon_1}).
In addition, the detected graviton frequency has now a peak at $60$ Hz and, thus, the energy of the graviton is $E_g= 2.48\times10^{-13}$ eV. Therefore for the
previously mentioned upper bound on the graviton mass, 
namely $1.24\times 10^{-22}\,{\rm eV}/c^2$, the difference between the speed of the photon and graviton becomes: 
$\D v_g  = \,c-v_g\, = \f{m_{g}^{2}c^5}{2E_{g}^{2}} \leq 3.75\times 10^{-11}$ m/sec. Fourth, the EM counterpart of  GW190521 was detected by the ZTF  \cite{Graham:2020gwr}. In particular, the ZTF ``sees" two frequency bands and specifically  for the GW190521, in the g-band the observed wavelength  is $\lambda_{\g g} = 4686\times 10^{-10}$ m while in the  r-band the observed wavelength is  $\lambda_{\g r}= 6166\times10^{-10}$ m. Therefore, the  energies corresponding to these observed photon wavelengths are $E_{\g g}=2.65$ eV and $E_{\g r}=2.01$ eV.
Finally, substituting the above numerical values in Eq. (\ref{bound_photon_1}), the upper bounds on the GUP parameter 
using the data for the r- and g- band  read, respectively,
\begin{eqnarray}
\beta_{0}\,\leq\,\frac{\D v_g\,M_{Pl}^{2}\,c^3}{3\,\left|\big(E_{g}^{2}-E_{\g r}^{2}\big)\right|}\,\bigg(1-\f{2GM}{r\,c^2}\bigg)=\, 1.54\times 10^{36}
\label{bound_photon_em_1}
\end{eqnarray}
\begin{eqnarray}
\beta_{0}\,\leq\,\frac{\D v_g\,M_{Pl}^{2}\,c^3}{3\,\left|\big(E_{g}^{2}-E_{\g g}^{2}\big)\right|}\,\bigg(1-\f{2GM}{r\,c^2}\bigg)=\, 8.83\times 10^{35}
\label{bound_photon_em_2}
\end{eqnarray}
%
%
where we have used the upper bound, obtained above, i.e. $\D v_g  =  3.75\times 10^{-11} \mbox{m/sec}$.
\\
The reader may note that in obtaining $\b_0$ as in the above equations, $E_g$ has negligible role in comparison to $E_{\g}$, as $E_{\g} \gg E_g$. This leads us to make a comment that the significant reduction in the upper bounds of $\b_0$ ( see Eqs. (\ref{bound_photon_em_1}) and (\ref{bound_photon_em_2})),  are solely due to the modifications emerging from the EM counterpart of the GW.

%
%
%
%
%
%
%
%
%
%
\section{Bound on the LQGUP parameter from GW150914}
\label{gup_linear}
%
%
%
%
\par\noindent
In this section, we will study the modifications in the difference between the speed of graviton and photon, i.e., $\D v_g$, in Schwarzschild spacetime while incorporating the LQGUP version. We define the 
physical position and momenta in terms of the canonical auxiliary variables as follows
%
%
%
%
\bseq
\begin{align}
&x_i=x_{0i},\,\,\,\,\,\,\,\,\,\,\,\,\,\,\,\,\,\,\,p_0=k_0
\label{gupmodel_2}\\[10pt]
& p_i=\,k_i\,(1-\a\,k+\,2\a^2\,k^2) ~.\label{gupmodel_2_a}
\end{align}
\eseq
We follow a procedure similar to Section III and allow terms up to the quadratic order of the parameter $\alpha$, as well as of the LQGUP parameter $\a_0$.  
Employing the expression for the GUP-modified squared 4-momentum and substituting  Eq. (\ref{gupmodel_2_a}), we obtain
\beq
g_{00}\,(p^{0})^{2}=\,-\,m_{g}^{2}c^2-\,p^2-\,2\a\,k k^2+\,5\a^2\,k^2 k^2~.
\label{mom_gup_2nd}
\eeq
We perform an inverse transformation of Eq. (\ref{gupmodel_2_a})  and write $k$ as a function of $p$ in the form
\beq
k^2=\,p^2\,(1+\,2\a\,k-\a^2\,k^2)~.
\label{eq_k_1}
\eeq
Solving the above quadratic equation, we get 
\beq
k=\,\f{\a p^2\,\pm\,\bigg[2\a^2\,p^2 p^2+\,p^2\bigg]^{1/2}}{(1+\,\a^2 \,p^2)}~
\label{root_k_1}
\eeq
which upon simplifying yields
\beq
k=  \a\,p^2   \pm\, p ~.
\label{root_k_2}
\eeq
The above, substituted in   
Eq. (\ref{mom_gup_2nd}), gives
\beq
(p^{0})^2=\,\f{1}{g_{00}}\bigg[\,-m_{g}^{2}c^2\,-p^2 \mp\,2\a\,p p^2\,-\a^2\,p^2 p^2\bigg]~.
\label{mom_gup_2nd_1}
\eeq
Following the analysis in Section III, we obtain  the energy of the graviton  to be
\beq
E_g=\,\sqrt{-g_{00}}\, \bigg[m_{g}^{2}c^4+\,p^2c^2\,(1\pm\a\,p)^2\bigg]^{1/2}
\label{energy_gup_2nd}
\eeq
which is the modified dispersion relation for
the LQGUP under consideration.
At this point, we follow the analysis as similar to Section III for the iteration method (see below Eq. (\ref {vel_gup_2})), and the  group velocity of the graviton becomes
\footnote{A full analysis, with no small $m_g$ approximation, is given in Appendix C.}
%
%
%
%
%
\beq
\tilde{v}_g=\,\f{1}{\sqrt{-\,g_{00}}}\,\f{\p E_g}{\p p}=\,c\,\bigg[1\pm\,\f{2\a \,E_g}{c}\bigg(1+\f{GM}{rc^2}\bigg)\bigg]~.
\label{vel_gup_2nd}
\eeq
%
%
%
%
%
%
%
%
%
%
Let us now consider separately the two cases in Eq. (\ref{vel_gup_2nd}) 
\beq
{\rm Case\,\,1\, :} \,\,\,\,\,\,\,\,\,\,\,\,\,&
\tilde{v}_{g}^{(1)}=\,c\,\bigg[1+\,\f{2\a \,E_g}{c}\bigg(1+\f{GM}{rc^2}\bigg)\bigg]\label{case_1_2nd}
\eeq
\beq
{\rm Case\,\,2\, :} \,\,\,\,\,\,\,\,\,\,\,\,\,&
\tilde{v}_{g}^{(2)}=\,c\,\bigg[1-\,\f{2\a \,E_g}{c}\bigg(1+\f{GM}{rc^2}\bigg)\bigg]~.
\label{case_2_2nd}
\eeq
%
%
%
\par\noindent
Eq. (\ref{case_1_2nd}) gives the speed of the graviton to be greater than $c$ which is the speed of photon in vacuum. 
%
We ignore such superluminal propagation and, thus, we proceed  with Case 2 associated with a subluminal graviton. 
\par\noindent
For this case, the difference between the speed of the graviton and the speed of the photon is now of the form
\beq
\D \tilde{v}_g=\,\left|c-\tilde{v}_g\right|=\,2\a\,E_g\,\bigg(1+\f{GM}{rc^2}\bigg)~.
\label{diff_vel_2nd_1}
\eeq
As in Section III, the difference in the speed of graviton and photon is bounded from above by $\D v_g$, and this sets a bound on the LQGUP parameters as follows
\beq
&&\a\,\leq\,\f{\D v_g}{2E_g}\,\bigg(1+\f{GM}{rc^2}\bigg)^{-1}\nonumber\\
&&\a_0\,\leq\,\f{\D v_g\,M_{Pl}c}{2\,E_g}\,\bigg(1-\f{GM}{rc^2}\bigg)
\label{bound_gup_2nd}
\eeq
where we have neglected terms of $O(\ell_{Pl}^4)$. 
%
%
%
%
%
%
%
%
%
%
%
%
%
%
%
Finally, if we use the data set from the event GW150914  \cite{Abbott:2016blz,TheLIGOScientific:2016src}, as we did in Section III, we obtain an  upper bound on the LQGUP parameter
\be
\a_0\, \leq  1.96\times 10^{20}~.
\ee
%
%
\par\noindent
At this point a couple of comments are in order. First, similar to the earlier cases, the contribution from the curvature of spacetime has no role in obtaining the bound on $\a_0$ and, thus, the upper bound on $\a_0$ turns out to be unmodified from that of the flat spacetime results, as obtained in Refs. \cite{Feng:2016tyt, Roushan:2019miz}.
Second, it is  expected that this bound on $\alpha_0$ will improve significantly over time with increasing accuracies of GW observations.

%
%
%
\subsection{$\D \tilde{v}_g$ with the LQGUP-modified speed of photon and GW190521}\label{vary_c_2}
%
%
%
%
%
\par\noindent
As in Subsection IIIA, we now consider the case that the velocity of the photon is also LQGUP-modified similar to the velocity of the graviton. Therefore, adopting the previous analysis for the velocity of the graviton with small mass approximation (see Eq. (\ref{vel_gup_2nd})), we get
\beq
\tilde{v}_{\g}=\,c\,\bigg[1\pm\,\f{2\a \,E_{\g}}{c}\bigg(1+\f{GM}{rc^2}\bigg)\bigg]~.
\label{vel_photon_2nd_1}
\eeq
We now explore all possible cases of $\tilde{v}_{\g}$ and $\tilde{v}_{g}$.
%
%
\beq
{\rm Case\,\,1\, :} \,\,\,\,\,\,\,\,\,\,\,\,\,&
\tilde{v}_{g}^{(1)}=\,c\,\bigg[1+\,\f{2\a \,E_g}{c}\bigg(1+\f{GM}{rc^2}\bigg)\bigg]\nonumber\\
&\,\,\,\,\, \tilde{v}_{\g}^{(1)}=\,c\,\bigg[1+\,\f{2\a \,E_{\g}}{c}\bigg(1+\f{GM}{rc^2}\bigg)\bigg]~.
\label{case_1_diff}
\eeq
%
%
%
%
%
%
%
%
%
%
%
%
%
%
%
%
%
%
%
%
%
%
%
Since the above equation gives the speed  of  the  graviton to  be  greater  than $c$, i.e., it is superluminal,  we drop this case from future considerations and proceed to the remaining Cases.
\beq
{\rm Case\,\,2\, :} \,\,\,\,\,\,\,\,\,\,\,\,\,&
\tilde{v}_{g}^{(2)}=\,c\,\bigg[1+\,\f{2\a \,E_g}{c}\bigg(1+\f{GM}{rc^2}\bigg)\bigg]\nonumber\\
&\,\,\,\,\, \tilde{v}_{\g}^{(2)}=\,c\,\bigg[1-\,\f{2\a \,E_{\g}}{c}\bigg(1+\f{GM}{rc^2}\bigg)\bigg]~.
\label{case_2_diff}
\eeq
%
%
%
%
%
%
%
%
%
%
%
%
%
%
%
%
%
%
As in Case 1, Case 2 also introduces a superluminal and hence unphysical graviton. Thus, we drop Case 2 as well. 
%
%
%
%
%
%
%
\beq
{\rm Case\,\,3\, :} \,\,\,\,\,\,\,\,\,\,\,\,\,&
\tilde{v}_{g}^{(3)}=\,c\,\bigg[1-\,\f{2\a \,E_g}{c}\bigg(1+\f{GM}{rc^2}\bigg)\bigg]\nonumber\\
&\,\,\,\,\, \tilde{v}_{\g}^{(3)}=\,c\,\bigg[1+\,\f{2\a \,E_{\g}}{c}\bigg(1+\f{GM}{rc^2}\bigg)\bigg]~.
\label{case_3_diff}
\eeq
%
%
%
%
%
%
%
%
%
%
%
%
%
%
%
%
%
%
%
%
%
In this case, the GUP-modified speed of photon is greater than c, i.e., it is superluminal, thus we drop this case from future considerations. 
\beq
{\rm Case\,\,4\, :} \,\,\,\,\,\,\,\,\,\,\,\,\,&
\tilde{v}_{g}^{(4)}=\,c\,\bigg[1-\,\f{2\a \,E_g}{c}\bigg(1+\f{GM}{rc^2}\bigg)\bigg]\nonumber\\
& \tilde{v}_{\g}^{(4)}=\,c\,\bigg[1-\,\f{2\a \,E_{\g}}{c}\bigg(1+\f{GM}{rc^2}\bigg)\bigg]~.
\eeq
%
%
The above equations gives us a difference between the speed of the graviton and photon of the form
%
%
\begin{eqnarray}
\Delta 
\tilde{v}_{g}^{(4)}=\,2\a\left|\big[E_{g}-E_{\g}\big]\right|\bigg(1+\f{GM}{rc^2}\bigg)
\label{diff_photon_5}
\end{eqnarray}
%
%
%
%
which leads to the bound on the LQGUP parameter of the form
%
%
%
%
\begin{eqnarray}
\alpha_0\,\leq\,\frac{\D v_{g}\,M_{Pl}c}{2\,\left|\big(E_{g}-E_{\g}\big)\right|}\,\bigg(1-\f{GM}{rc^2}\bigg)~.
\label{bound_photon_5}
\end{eqnarray}
Finally, we utilize the data for the event GW190521 as  given by  LIGO Scientific and Virgo  Collaborations \cite{Abbott:2020tfl} and by ZTF \cite{Graham:2020gwr}. Using the data for the r- and g- bands, we obtain  the upper bound on the LQGUP parameter, respectively,  as follows
\be
\a_0\,\leq\,\f{\D v_{g}\,M_{Pl}c}{2\,\left|\big(E_{g}-E_{\g r}\big)\right|}\,\bigg(1-\f{GM}{rc^2}\bigg)\,=3.79\times 10^{8}
\label{alpha_em_1}
\ee
and
\be
\a_0\,\leq\,\f{\D v_{g}\,M_{Pl}c}{2\,\left|\big(E_{g}-E_{\g g}\big)\right|}\,\bigg(1-\f{GM}{rc^2}\bigg)\,=2.88\times 10^{8}
\label{alpha_em_2}
\ee
where we use the upper bound $\D v_g  = 3.75\times 10^{-11}$ m/sec. It should be noted that also here the curvature of the spacetime has practically  no role in obtaining the upper bound on $\a_0$. Furthermore, we emphasize that in obtaining $\a_0$ as in the above equations, $E_{\g}$ dominates over $E_g$, as $E_{\g} \gg E_g$. This indicates that 
the significant reduction in the upper bounds of $\a_0$ (see Eqs. (\ref{alpha_em_1}) and (\ref{alpha_em_2})),  are solely due to the modifications as emerging from the EM counterpart of the GW.
%
%
%
%
%
%
\section{Conclusion}\label{conclusions}
\par\noindent
The existence of a minimum measurable length has been predicted by candidate theories of QG as well as from other considerations such as from the physics of black holes. This necessitates the modification of 
HUP to GUP. An implication of GUP is the modification of
the standard dispersion relation and, consequently, of the speed of particles. From theoretical studies on GWs, one can bound the mass of the graviton and, thus, the speed of the graviton. 
In our analysis, we consider two versions of GUP: one with a quadratic term in momentum (GUP) and the other with linear and  quadratic terms in momentum (LQGUP). 
\par\noindent
In the current work, we consider a curved spacetime background and employing GUP, we obtain upper bounds for the GUP parameters.
Using the data from GW150914, the GUP parameter is bounded as $\beta_{0} < 2.56\times 10^{60}$ while the LQGUP parameter is bounded as $\alpha_{0} < 1.96\times 10^{20}$. From these results, it is evident that the effects of the curved background are negligible since our results are the same with the existing ones in the literature, which were derived for a flat background. 
However, one can consider  the speed of the photon to be modified as well, along with the speed of the graviton. In this case, one can use the data from GW190821, since this is the only gravitational event between two black holes which has electromagnetic part that was detected by ZTF. So taking into consideration the observed energies of the photons, the GUP parameter is bounded as $\beta_{0} < 1.54\times 10^{36}$ for the r-band observed by ZTF, and $\beta_{0} < 8.83\times 10^{35}$ for the g-band observed by ZTF.
The LQGUP parameter is bounded as $\alpha_{0} < 3.79\times 10^{8}$ for the r-band observed by ZTF, and $\alpha_{0} < 2.88\times 10^{8}$ for the g-band observed by ZTF.
These bounds are among the best 
compared to the ones existing in the literature, and specifically the bounds on the dimensionless LQGUP  parameter $\alpha_{0}$ are the tightest. Therefore, GW observations give strict bounds on the GUP parameters  \footnote{It is noteworthy that this strict bounds on the GUP parameters could be extended to the polymer quantization parameter. This is due to the fact that the polymer quantization   provides modified uncertainty relations similar to the ones of GUP \cite{Hossain:2010wy,Majumder:2012qy}.}, especially when employing data from different ``messenger"  signals which describe the same gravitational event. This underscores the need for 
more and further advanced gravitational wave detectors and, that of multimessenger observations.
We note that in our analysis, we have assumed small but non-zero masses for the graviton and photon, 
consistent with observational bounds on these masses. 
\par\noindent
Finally, it should be noted that recently, the LIGO Scientific and Virgo Collaborations have released their updated catalogue of GW detections, i.e., GWTC-2, in which the bound on the mass of the graviton is even stricter and, for the first time, based on observational data. The new upper bound on the graviton mass is 
$m_g \leq 1.76\times 10^{-23}$  eV/c$^2$ \cite{Abbott:2020jks}. 
This bound is tighter by a factor of 10 which means that all the bounds derived in our work can be improved by a factor of 10.\\
%
%
%
%
%
%
%
%
%
%
\section{Acknowledgments} 
%
%
%
\par\noindent
ECV would like to thank E. Berti and M. Kasliwal for useful correspondences and fruitful comments.
This work was supported by the Natural Sciences and Engineering Research Council of Canada. 
%
%
%
%
%
%
%
%
\appendix
%
%
%

\section{Speed of photon and graviton in the Schwarzschild background with massive photon}\label{app_1}
\par\noindent
We consider the photon with a small but non-zero mass. Then, following the analysis in Section II and employing  Eq. (\ref{vel_gr_4}), adapted to the massive photon, we obtain
\beq
v_{\g}=\,c\bigg[1-\f{m_{\g}^{2}c^4}{2E_{\g}^{2}}\,\bigg(1-\f{2GM}{rc^2}\bigg)\bigg]
\label{vel_ph_mas_1}
\eeq
where $m_{\g}$ is the mass of the photon and we have reasonably assumed $\f{m_{\g}^{2}c^4}{2E_{\g}^{2}}\,\ll1$. Therefore, the difference in the speed of the graviton and photon assumes the form
\beq
\D v_g=\, \f{c^5}{2}\bigg(1-\f{2GM}{rc^2}\bigg)\bigg(\f{m_{g}^{2}}{E_{g}^{2}}-\f{m_{\g}^{2}}{E_{\g}^{2}}\bigg)~.
\label{diff_vel_mass_1}
\eeq
%
%
%
%
\section{Bound on GUP parameter with massive photon and graviton}\label{app_2}
%
%
%
%
%
\par\noindent
One may now want to include QG effects due to the GUP version, 
as we did in Section III, when the photon and graviton are both massive. So, we get the zeroth order solution for $p$ by setting $\b =0$ in Eq. (\ref{energy_gr_1}) which leads to $p=\pm\,\bigg[\f{E_{g}^{2}}{(-g_{00})c^2}-m_{g}^{2}c^2\bigg]^{1/2}$ and  consider only the positive root for the momentum. For our convenience, we define $\mathcal{E}_g = \bigg[\f{E_{g}^{2}}{(-g_{00})c^2}-m_{g}^{2}c^2\bigg]$ and substitute it in Eq. (\ref{energy_gr_1}) in order to get the momentum in the form $p=p(E_g)$. 
Finally, we substitute in the (LHS) of Eq. (\ref{energy_gr_1}) the zeroth order expression  $\f{E_g}{\sqrt{-g_{00}}}=(m_{g}^{2}c^4+p^2c^2)^{1/2}$ and get
\beq
(m_{g}^{2}c^4+p^2c^2)^{1/2}=\,\bigg[m_{g}^{2}c^4+\,c^2\mathcal{E}_g\,(1-2\beta \mathcal{E}_g)\bigg]^{1/2}
\label{dis_mas_1}
\eeq
which yields
\beq
&&p^2=\,\mathcal{E}_g\,(1-2\b\,\mathcal{E}_g)
\label{dis_mas_2}\\
&&p=\,\mathcal{E}^{1/2}_g\,(1-\beta\,\mathcal{E}_g+{\cal O}(\b^2))~.
\label{dis_mas_3}
\eeq
Since $\beta \sim \ell_{Pl}^{2}$, we neglect terms of higher order of  $\b$, so Eq. (\ref{dis_mas_3}) becomes $p(E_g)=\mathcal{E}_g^{1/2}\,(1-\b\,\mathcal{E}_g)$ and substituting it in Eq. (\ref{vel_gup_1}), we get 
\beq
\tilde{v}_g&=&\,\f{p c^2\,(1-4\b\,p^{2})}{\bigg[m_{g}^{2}c^4+p^{2}c^2-2\b\, p^{2}\,p^{2}c^2\bigg]^{1/2}}\nonumber\\
&=&\,c(1-4\b\,p^{2})\,\bigg[1+\bigg(\f{m_{g}^{2}c^2}{p^{2}}-2\b\,p^{2}\bigg)\bigg]^{-1/2}~.
\label{vel_mass}
\eeq
Now, we expand, the last term in the (RHS) of Eq. (\ref{vel_mass}),  which is  in the square brackets  with respect to the parameter $\b$ and neglecting the terms of higher order of  $\b$,  we obtain
\beq
\tilde{v}_g&=&\,c(1-4\b\,p^{2})\,\bigg[1-\f{m_{g}^{2}c^2}{2p^{2}}+\b\, p^{2}+\f{3m_{g}^{4}c^4}{8p^{4}}-\f{3\,\b m_{g}^{2}c^2}{2}\bigg]\nonumber\\
&=&\,c\bigg[1-\f{m_{g}^{2}c^2}{2\mathcal{E}_{g}}-\f{\b\,m_{g}^{2}c^2}{2}+\f{3\,m_{g}^{4}c^4}{8\mathcal{E}_{g}^2}-3\b\,\mathcal{E}_{g}\bigg] ~.
\label{vel_mass_1}
\eeq
We evaluate the second term in the (RHS) of Eq. (\ref{vel_mass_1}) 
\beq
\f{m_{g}^{2}c^2}{2\mathcal{E}_{g}}=\,\f{m_{g}^{2}c^2}{2\bigg[\f{E_{g}^{2}}{c^2\,(-g_{00})}-m_{g}^{2}c^2\bigg]}=\f{m_{g}^{2}c^2}{2\bigg[\f{E_{g}^{2}}{c^2\,\bigg(1-\f{2GM}{rc^2}\bigg)}-m_{g}^{2}c^2\bigg]}=\f{m_{g}^{2}c^2}{2\bigg[\f{E_{g}^{2}}{c^2}\,\bigg(1+\f{2GM}{rc^2}\bigg)-m_{g}^{2}c^2\bigg]}
\eeq
in which, using the data from GW190521, the terms below assume the values

\beq
\f{2GM}{rc^2}\sim \, 0.025\times 10^{-19},\,\,\,\,\,\,\,\,\,\,\,\,\,\,\,m_g\sim\,1.24\times 10^{-22}{\rm eV}/c^2,\,\,\,\,\,\,\,\,\,\,\,\,\,\,E_g=2.49\times10^{-13}{\rm eV}~.
\eeq
So, utilizing the above numerical values the second and fourth terms in the (RHS) of Eq. (\ref{vel_mass_1}) become $\f{m_{g}^{2}c^2}{2\mathcal{E}_{g}} \sim\, 1.24\times 10^{-19}$ and \, $\f{3m_{g}^{4}c^4}{8\mathcal{E}_{g}^2}\sim \, 2.3\times 10^{-38}$.
Therefore, neglecting  the terms  $\f{m_{g}^{2}c^2}{2\mathcal{E}_{g}}$ and  $\f{3m_{g}^{4}c^4}{8\mathcal{E}_{g}^2}$ with respect to unity in Eq. (\ref{vel_mass_1}),
%
%
%
%
%
%
%
the speed of the graviton  reads
\beq
\tilde{v}_g=\,c-\f{\b\,m_{g}^{2}c^3}{2}-3\b\,\mathcal{E}_{g}c~.
\label{vel_mass_grav_1}
\eeq
Following exactly the same analysis for the case of the massive photon, the speed of the photon will be of the form
\beq
\tilde{v}_{\g}=\,c-\f{\b\,m_{\g}^{2}c^3}{2}-3\b\,\mathcal{E}_{\g}c~.
\label{vel_mass_ph_1}
\eeq
At this point a number of comments are in order.
First,  we use the data for the event GW190521, and, specifically, for the photon energy, we employ the numerical values given by ZTF, thus $E_{\g g}=2.65$ eV and $E_{\g r}=2.02$ eV. It is clear that these numerical values  are much higher than the energy of the graviton, i.e., $E_{g}$, and, thus, the corresponding quantity $\mathcal{E}_{\gamma}$ is larger than    $\mathcal{E}_{g}$. Second, taking the above-mentioned comments into consideration, similar to the massive graviton,  we reasonably neglect the terms $\f{m_{\g}^{2}c^2}{2\mathcal{E}_{\g}}$ and  $\f{3m_{\g}^{4}c^4}{8\mathcal{E}_{\g}^{2}}$ with respect to unity. 
\par\noindent
Therefore,  the difference in the speed of the graviton and photon now reads
\beq
\D \tilde{v}_g=\left|\tilde{v}_{\g}-\tilde{v}_g\right|=\, \b\bigg[\f{5c^3(m_{\g}^{2}-m_{g}^{2})}{2}+\f{3}{c}\bigg(1+\f{2GM}{rc^2}\bigg)\left|(E_{g}^{2}-E_{\g}^{2})\right|\bigg]~.
\label{diff_gr_mass_1}
\eeq
Therefore, as in Section III, the difference in the speed of the graviton and photon is bounded from above by $\D v_g$ as given by Eq. (\ref{diff_vel_mass_1}), and this sets a bound on the GUP parameter as follows
\beq
&\b\,\bigg[\f{5c^3(m_{\g}^{2}-m_{g}^{2})}{2}+\f{3}{c}\bigg(1+\f{2GM}{rc^2}\bigg)\left|(E_{g}^{2}-E_{\g}^{2})\right|\bigg]\,\leq\,\D v_g\nonumber\\
\implies &\b_0\,\leq\,\D v_g\, M_{Pl}^{2}c^2\,\bigg[\f{5c^3(m_{\g}^{2}-m_{g}^{2})}{2}+\f{3}{c}\bigg(1+\f{2GM}{rc^2}\bigg)\left|(E_{g}^{2}-E_{\g}^{2})\right|\bigg]^{-1}~.
\label{bound_mass_I}
\eeq 
A couple of comments are in order. First, {if we had a bound for the mass of the photon in the context of GW observations, then from Eq. (\ref{bound_mass_I}) the upper bound on $\beta_0$ can be estimated. This is because all other quantities are known for the event GW190521 by the data given for the gravitons by the LIGO Scientific and Virgo  Collaborations \cite{Abbott:2020tfl} and for the photons by ZTF \cite{Graham:2020gwr}. Second, if the masses of graviton and photon are taken to be zero, i.e., 
$(m_{\g},m_g)\,\to 0$, one obtains Eq. (\ref{diff_photon_1}) from Eq.(\ref{diff_gr_mass_1}), as expected.
%
%
\section{Bound on LQGUP parameter with massive photon and graviton}
%
%
%
%
%
\par\noindent
One may now want to include QG effects due to LQGUP version, as we did in Section IV, when the photon and graviton are both massive. So, we get the zeroth order solution for the momentum $p$ of Eq. (\ref{energy_gup_2nd}) without the small mass approximation for the graviton. 
For our convenience, we use again the quantity $\mathcal{E}_g = \bigg[\f{E_{g}^{2}}{(-g_{00})c^2}-m_{g}^{2}c^2\bigg]$ and substitute it in Eq. (\ref{energy_gup_2nd}) in order to get the momentum in the form $p=p(E_g)$. Then, we follow a similar analysis as in Appendix B to get
%
%
%
%
%
%
%
%
%
%
%
%
%
%
\beq
p^{2}=\,\mathcal{E}_g\,(1\pm\,\a\,\mathcal{E}_{g}^{1/2})^2
\label{mom_mass_gupII}
\eeq
and the speed of the graviton now reads
\beq
\tilde{v}_g=\,\f{1}{\sqrt{-g_{00}}}\f{\p E_g}{\p p}&=&\,\f{\p}{\p p}\bigg[m_{g}^{2}c^4+p^{2}c^2\,(1\pm\a\,p)^2\bigg]^{1/2}\nonumber\\
&=&\f{p c^2+\,2\a^2c^2\,p p^{2}\pm\,3\a c^2 p^{2}}{\bigg[m_{g}^{2}c^4+\,p^{2}c^2\,(1\pm\,2\a\,p+\,\a^2p^{2})\bigg]^{1/2}}~.
\label{vel_mass_gr_2}
\eeq
\par\noindent
We now expand the denominator in the (RHS) of Eq. (\ref{vel_mass_gr_2}) and keeping terms up to $\mathcal{O}(\a^2)$, we get
\beq
&&\bigg[m_{g}^{2}c^4+\,p^{2}c^2\,(1\pm\,2\a\,p+\,\a^2p^{2})\bigg]^{-1/2}= (pc)^{-1}\,\bigg[1+\bigg(\f{m_{g}^{2}c^2}{p^{2}}\pm 2\a\,p+\,\a^2\,p^{2}\bigg)\bigg]^{-1/2}\\
&=&(pc)^{-1}\,\bigg[1-\f{1}{2}\bigg(\f{m_{g}^{2}c^2}{p^{2}}\pm 2\a\,p+\,\a^2\,p^{2}\bigg)+\f{3}{8}\bigg(\f{m_{g}^{2}c^2}{p^{2}}\pm 2\a\,p+\,\a^2\,p^{2}\bigg)^2+ \dots\bigg]\\
&=&(pc)^{-1}\,\bigg[1-\f{m_{g}^{2}c^2}{2p^{2}}+\f{3\,m_{g}^{4}c^4}{8p^{4}}\mp \a\,p\pm \,\f{3\a\,m_{g}^{2}c^2}{2 p}+\,\a^2\,p^{2}+\,\f{3\a^2\,m_{g}^{2}c^2}{4}\bigg]
\label{denomin_mass_1}~.
\eeq
It is easily seen that there are two cases to be studied in Eq. (\ref{vel_mass_gr_2}), one for each sign. 
Therefore, we take the first case and write the Eq. (\ref{vel_mass_gr_2}) as follows
%
%
%
%
%
\beq
\tilde{v}_g&=&\,c\,\bigg[1+\,2\a^2\,p^{2}+\,3\a\,p\bigg]\,\bigg[1-\f{m_{g}^{2}c^2}{2p^{2}}+\f{3\,m_{g}^{4}c^4}{8p^{4}}-\a\,p+\,\f{3\a\,m_{g}^{2}c^2}{2 p}+\,\a^2\,p^{2}+\,\f{3\a^2\,m_{g}^{2}c^2}{4}\bigg]\\
&=&\,c\,\bigg[1-\,\f{m_{g}^{2}c^2}{2p^{2}}+\,\f{3\,m_{g}^{4}c^4}{8p^{4}}+\,2\a\,p+\,\f{9\a\,m_{g}^{4}c^4}{8\,p^{3}}+\f{17\a^2\,m_{g}^{2}c^2}{4}+\f{3\a^2\,m_{g}^{4}c^4}{4\,p^{2}}\bigg]~.
\eeq
Now, we write the above equation in terms of $\mathcal{E}_g$ and utilize the same analysis as the one adopted in the previous Appendix. We neglect the terms $\f{m_{g}^{2}c^2}{2\mathcal{E}_g}$ and  $\f{3m_{g}^{4}c^4}{8\mathcal{E}_{g}^{2}}$ with respect to unity and so the speed of the graviton becomes
\beq
\tilde{v}_g=\,c\,\bigg[1+\,2\a\,\mathcal{E}^{1/2}_{g}+\,\f{\a\, m_{g}^{2}c^2}{\mathcal{E}_{g}^{1/2}}-\,\f{3\a\, m_{g}^{4}c^4}{8\,\mathcal{E}_{g}^{3/2}}+\,2\a^2\,\mathcal{E}_{g}+\,
\f{9\a^2\,m_{g}^{4}c^4}{8\,\mathcal{E}_{g}}
+\f{11\a^2\,m_{g}^{2}c^2}{4}\,\bigg]~.
\label{vel_mass_gr_3}
\eeq
Similarly, the speed of photon becomes
\beq
\tilde{v}_{\g}=\,c\,\bigg[1+\,2\a\,\mathcal{E}^{1/2}_{\g}+\,\f{\a\, m_{\g}^{2}c^2}{\mathcal{E}_{\g}^{1/2}}-\,\f{3\a\, m_{\g}^{4}c^4}{8\,\mathcal{E}_{\g}^{3/2}}+\,2\a^2\,\mathcal{E}_{\g}+\,
\f{9\a^2\,m_{\g}^{4}c^4}{8\,\mathcal{E}_{\g}}
+\f{11\a^2\,m_{\g}^{2}c^2}{4}\,\bigg]~.
\label{vel_mass_ph_2}
\eeq
\par\noindent
Therefore, the difference between the speed of graviton and photon turns out to be
\beq
\left|\D \tilde{v}_{g}\right|=\left|\tilde{v}_{\g}-\tilde{v}_{g}\right|=\,\Bigg|\bigg[2\a c\,(\mathcal{E}_{\g}^{1/2}-\mathcal{E}_{g}^{1/2})&+&\a\,c^3\bigg(\f{m_{\g}^{2}}{\mathcal{E}_{\g}^{1/2}}-\f{m_{g}^{2}}{\mathcal{E}_{g}^{1/2}}\bigg)-\,\f{3\a\, c^5}{8}\bigg(\f{m_{\g}^{4}}{\mathcal{E}_{\g}^{3/2}}-\f{m_{g}^{4}}{\mathcal{E}_{g}^{3/2}}\bigg)+\,2\a^2c(\mathcal{E}_{\g}-\mathcal{E}_{g})+\nonumber\\
&&\f{9\a^2c^5}{8}\bigg(\f{m_{\g}^{4}}{\mathcal{E}_{\g}}-\f{m_{g}^{4}}{\mathcal{E}_g}\bigg)+\f{11\a^2\,c^3}{4}\bigg(m_{\g}^{2}-m_{g}^{2}\bigg)\bigg]\Bigg|
\label{vel_mass_ph_3}~.
\eeq
As in Section IV, the difference in the speed of graviton and photon is bounded from above by $\D v_g$ as given by Eq. (\ref{diff_vel_mass_1}), and this sets a bound on the LQGUP parameter as follows
\beq
\bigg[2\a c\,(\mathcal{E}_{\g}^{1/2}-\mathcal{E}_{g}^{1/2})&+&\a\,c^3\bigg(\f{m_{\g}^{2}}{\mathcal{E}_{\g}^{1/2}}-\f{m_{g}^{2}}{\mathcal{E}_{g}^{1/2}}\bigg)-\,\f{3\a\, c^5}{8}\bigg(\f{m_{\g}^{4}}{\mathcal{E}_{\g}^{3/2}}-\f{m_{g}^{4}}{\mathcal{E}_{g}^{3/2}}\bigg)+\,2\a^2c(\mathcal{E}_{\g}-\mathcal{E}_{g})+\nonumber\\
&&\f{9\a^2c^5}{8}\bigg(\f{m_{\g}^{4}}{\mathcal{E}_{\g}}-\f{m_{g}^{4}}{\mathcal{E}_g}\bigg)+\f{11\a^2\,c^3}{4}\bigg(m_{\g}^{2}-m_{g}^{2}\bigg)\bigg]\,\leq\,\D v_g~.
\label{bound_mass_1}
\eeq
It should be noted that both particles, namely graviton and photon, in this case appear to be superluminal particles.
Next,  we take the second possible case of Eq. (\ref{vel_mass_gr_2}) which leads to the speed of the graviton to be of the form
\beq
\tilde{v}_g&=&\,c\,\bigg[1-\,3\a\,p+\,2\a^2\,p^{2}\bigg]\,\bigg[1-\f{m_{g}^{2}c^2}{2p^{2}}+\f{3\,m_{g}^{4}c^4}{8p^{4}}+\a\,p-\,\f{3\a\,m_{g}^{2}c^2}{2 p}+\,\a^2\,p^{2}+\,\f{3\a^2\,m_{g}^{2}c^2}{4}\bigg]\\
&=&\,c\,\bigg[1-\,\f{m_{g}^{2}c^2}{2p^{2}}+\,\f{3\,m_{g}^{4}c^4}{8p^{4}}-\,2\a\,p-\,\f{9\a\,m_{g}^{4}c^4}{8\,p^{3}}+\f{17\a^2\,m_{g}^{2}c^2}{4}+\f{3\a^2\,m_{g}^{4}c^4}{4\,p^{2}}\bigg]~.
\eeq
Employing now $p^{2}=\,\mathcal{E}_g\,(1-\,\a\,\mathcal{E}_{g}^{1/2})^2$ in the above equation and proceeding as before,  we obtain
\beq
\tilde{v}_g=\,c\,\bigg[1-\,2\a\,\mathcal{E}^{1/2}_{g}-\,\f{\a\, m_{g}^{2}c^2}{\mathcal{E}_{g}^{1/2}}+\,\f{3\a\, m_{g}^{4}c^4}{8\,\mathcal{E}_{g}^{3/2}}+\,2\a^2\,\mathcal{E}_{g}+\,
\f{9\a^2\,m_{g}^{4}c^4}{8\,\mathcal{E}_{g}}
+\f{11\a^2\,m_{g}^{2}c^2}{4}\,\bigg]
\label{vel_mass_gr_2_2}
\eeq
and
\beq
\tilde{v}_{\g}=\,c\,\bigg[1-\,2\a\,\mathcal{E}^{1/2}_{\g}-\,\f{\a\, m_{\g}^{2}c^2}{\mathcal{E}_{\g}^{1/2}}+\,\f{3\a\, m_{\g}^{4}c^4}{8\,\mathcal{E}_{\g}^{3/2}}+\,2\a^2\,\mathcal{E}_{\g}+\,
\f{9\a^2\,m_{\g}^{4}c^4}{8\,\mathcal{E}_{\g}}
+\f{11\a^2\,m_{\g}^{2}c^2}{4}\,\bigg]
\label{vel_mass_ph_2_2}~.
\eeq
In this second case, the difference between the speed of graviton and photon reads
\beq
\left|\D \tilde{v}_{g}\right|=\left|\tilde{v}_{\g}-\tilde{v}_{g}\right|=\,\Bigg|\bigg[\a\,c^3\bigg(\f{m_{\g}^{2}}{\mathcal{E}_{\g}^{1/2}}-\f{m_{g}^{2}}{\mathcal{E}_{g}^{1/2}}\bigg)&-&2\a c\,(\mathcal{E}_{\g}^{1/2}-\mathcal{E}_{g}^{1/2})+\,\f{3\a\, c^5}{8}\bigg(\f{m_{\g}^{4}}{\mathcal{E}_{\g}^{3/2}}-\f{m_{g}^{4}}{\mathcal{E}_{g}^{3/2}}\bigg)+\,2\a^2c(\mathcal{E}_{\g}-\mathcal{E}_{g})+\nonumber\\
&&\f{9\a^2c^5}{8}\bigg(\f{m_{\g}^{4}}{\mathcal{E}_{\g}}-\f{m_{g}^{4}}{\mathcal{E}_g}\bigg)+\f{11\a^2\,c^3}{4}\bigg(m_{\g}^{2}-m_{g}^{2}\bigg)\bigg]\Bigg|
\label{vel_mass_ph_3_2}~.
\eeq
As stated before, the difference in the speed of graviton and photon is bounded from above by $\D v_g$ as given by Eq. (\ref{diff_vel_mass_1}), and this sets a bound on the LQGUP parameter as follows
\beq
\Bigg|\bigg[\a\,c^3\bigg(\f{m_{\g}^{2}}{\mathcal{E}_{\g}^{1/2}}-\f{m_{g}^{2}}{\mathcal{E}_{g}^{1/2}}\bigg)&-&2\a c\,(\mathcal{E}_{\g}^{1/2}-\mathcal{E}_{g}^{1/2})+\,\f{3\a\, c^5}{8}\bigg(\f{m_{\g}^{4}}{\mathcal{E}_{\g}^{3/2}}-\f{m_{g}^{4}}{\mathcal{E}_{g}^{3/2}}\bigg)+\,2\a^2c(\mathcal{E}_{\g}-\mathcal{E}_{g})+\nonumber\\
&&\f{9\a^2c^5}{8}\bigg(\f{m_{\g}^{4}}{\mathcal{E}_{\g}}-\f{m_{g}^{4}}{\mathcal{E}_g}\bigg)+\f{11\a^2\,c^3}{4}\bigg(m_{\g}^{2}-m_{g}^{2}\bigg)\bigg]\Bigg|\,\leq\,\D v_g~.
\eeq


%
%
%

\end{document}